\documentclass[11pt]{article}
\usepackage[margin=1in]{geometry}
\usepackage{amsmath,slashed,feynmf,epsfig,euscript,url,float,booktabs,verbatim}
\usepackage{bm}

\def\beq{\begin{equation}}
\def\eeq#1{\label{#1}\end{equation}}
\def\eeqn{\end{equation}}

\def\beqa{\begin{eqnarray}}
\def\eeqa#1{\label{#1}\end{eqnarray}}
\def\eeqan{\end{eqnarray}}

\let\bar=\overbar

\def\Dslash{\not{\hbox{\kern-4pt $D$}}}
\def\dslash{\not{\hbox{\kern-2pt $\del$}}}

\def\msb{{\bar{\ssstyle M \kern -1pt S}}}

\def\Title#1{\begin{center} {\Large {\bf #1} } \end{center}}
\def\Author#1{\begin{center} {\normalsize {\sc #1} } \end{center}}
\def\Institution#1{\begin{center} {\normalsize {\it #1} } \end{center}}
\def\Abstract#1{\noindent {\normalsize {\bf Abstract:} {\normalfont #1}}}
\def\Conference{\vspace{4mm}\begin{raggedright} {\normalsize {\it Talk presented at the 2019 Meeting of the Division of Particles and Fields of the American Physical Society (DPF2019), July 29--August 2, 2019, Northeastern University, Boston, C1907293.} } \end{raggedright}\vspace{4mm}}

\begin{document}

%
%

\Title{Reevaluating Uncertainties in $\bar{B}\rightarrow X_s\gamma$}

\Author{Ayesh Gunawardana}

\Institution{Department of Physics and Astronomy\\ Wayne State University, Detroit, Michigan 48201, USA}

\Abstract{The rare inclusive decay $\bar{B}\rightarrow X_s\gamma$ is an important probe of physics beyond the standard model. The largest uncertainty on the decay rate and CP asymmetry comes from operators that appear at order $1/m_b$ in the heavy quark expansion. One of the three leading contributions in the heavy quark expansion, $Q_1^q-Q_{7\gamma}$ is described by a non-local function whose moments are related to HQET parameters. We use recent progress in our knowledge of these parameters to better constrain their contributions to the total rate and CP asymmetry. }

\Conference

%
%

\section{Introduction}

The inclusive radiative $\bar{B}\rightarrow X_s\gamma$ decay is an important new physics probe. Since this is a flavor changing neutral current (FCNC) process, it does not occur at tree level in the standard model (SM). Therefore, it can be highly sensitive to new physics. In particular, these new physics sources can modify the Wilson coefficient $C_{7\gamma}$, and they may introduce new weak phases that can enhance the SM CP asymmetry.\par
At low energies the decay is described by the operator $Q_{7 \gamma}=\left(-e / 8 \pi^{2}\right) m_{b} \overline{s} \sigma_{\mu \nu} F^{\mu \nu}\left(1+\gamma_{5}\right) b$. This operator describes the coupling of a photon to the $b\rightarrow s$ weak vertex. However, this is not the only way to produce a photon. For instance, a gluon or a quark pair that was produced at the weak vertex can be converted to a photon, and these processes are described by the operator $Q_{8 g}=\left(-g / 8 \pi^{2}\right) m_{b} \overline{s} \sigma_{\mu \nu} G^{\mu \nu}\left(1+\gamma_{5}\right) b$ and $Q_{1}^{c}=(\overline{c} b)_{V-A}(\overline{s} c)_{V-A}$ respectively. Such a conversion will ``cost" a factor of $\alpha_s$ or $\frac{\Lambda_{\text{QCD}}}{m_b}$.\par  
Altogether, all these operators are considered in the full effective Hamiltonian for the decay,
\begin{eqnarray}\label{EffHamiltonian}
\mathcal{H}_{\mathrm{eff}}=\frac{G_{F}}{\sqrt{2}} \sum_{q=u, c} V_{q b}^{*} V_{q s}\left(C_{1} Q_{1}^{q}+\sum_{i=2}^{6} C_{i} Q_{i}+C_{7 \gamma} Q_{7 \gamma}+C_{8 g} Q_{8 g}\right)+\mathrm{h.c.}
\end{eqnarray}
where the $C_i$ 's are the Wilson coefficients, which contain the short distant physics. For our analysis $C_1, C_{7\gamma}$ and $C_{8g}$ are the most important Wilson coefficients since they are large compared to the rest.\par
At the leading order the decay rate is described by operator pair $Q_{7 \gamma}-Q_{7 \gamma}$. The largest uncertainty of the decay rate comes from the resolved photon contributions. They first appear at the order $1/m_b$ in the \textit{heavy quark expansion} and arise from operator pairs $Q_{1}-Q_{7 \gamma}, Q_{8 g}-Q_{8 g} \text { and } Q_{7 \gamma}-Q_{8 g}$.\par 
The experimental world average for the branching fraction is $\mathcal{B}\left(B \rightarrow X_{s} \gamma\right)\left(E_{\gamma}>1.6 \mathrm{\, GeV}\right)=(3.32 \pm 0.15) \times 10^{-4}$   
\cite{Tanabashi:2018oca}. The SM NNLO prediction is $\mathcal{B}_{s \gamma}^{\mathrm{SM}}=(3.36 \pm 0.23) \times 10^{-4}$ for the photon energy cut $E_{\gamma}>1.6 \mathrm{\,GeV}$ \cite{Misiak:2015xwa}. This prediction is obtained by using 
\begin{eqnarray}\label{NNLOprediction}
\Gamma\left(\overline{B} \rightarrow X_{s} \gamma\right)=\underbrace{\Gamma\left(b \rightarrow X_s \gamma\right)}_{\text {Perturbatively calculable }}+\underbrace{\delta \Gamma_{\text {nonp }}}_{O\left(\frac{\Lambda_{\text{QCD}}}{m_{b}}\right)}
\end{eqnarray}
Here $\Gamma\left(b \rightarrow X_{s} \gamma\right)$ term represents the decay of the constituent $b$ quark into a charmless final state \cite{Misiak:2015xwa}. The $\delta \Gamma_{\mathrm{nonp}}$ provides the contribution from the resolved photon effects, and they induce $\sim 5\%$ uncertainty to the total inclusive decay rate \cite{Benzke:2010js}. \par
The schematic form of the non-perturbative contribution is given as 
\begin{eqnarray}
\Delta \Gamma \sim \quad \underbrace{J}_{\text {Perturbatively calculable }} \otimes \underbrace{h}_{\text {Non perturbative }}.
\end{eqnarray}
The $J$ functions are the perturbatively calculable part. The soft $h$ functions are Fourier transforms of non local matrix elements that encode the long distance non-perturbative effects. In \cite{Benzke:2010js} the estimates for these non-perturbative contributions to the error were given as $ Q_{1}^{c}-Q_{7 \gamma} \in[-1.7,+4.0] \%$ and $Q_{8 g}-Q_{8 g} \in[-0.3,+1.9] \%$. For the $Q_{7 \gamma}-Q_{8 g}$ contribution two values were given. One was obtained from the vacuum insertion approximation (VIA), $[-2.8,-0.3] \%$ and other based on 2010 experimental data, $[-4.4,+5.6] \%$. The later is related to $\Delta_{0-}$, the isospin asymmetry of the neutral and charged $B$ decay to $X_s\gamma$. The new \textit{Belle} estimate and PDG average for $\Delta_{0-}$ \cite{Tanabashi:2018oca,Watanuki:2018xxg} reduce the $Q_{7\gamma}-Q_{8g}$ contribution to $[-1.4,+2] \%$ \cite{Gunawardana:2019gep} . Therefore, currently $Q_{1}^q-Q_{7\gamma}$ is the largest contributor to the error. In the following we would like to explore the possibility of reducing the size of the $Q_{1}^q-Q_{7\gamma}$ contribution.\par
Specifically the contribution to the non-perturbative uncertainty  from $Q_{1}^q-Q_{7\gamma}$ is given by:
\begin{eqnarray}
\mathcal{F}_E|_{17}=\frac{C_{1}}{C_{7 \gamma}} \frac{\Lambda_{17}}{m_{b}},
\end{eqnarray}
where 
\begin{eqnarray}\label{Lamda17}
\Lambda_{17}=e_{c} \operatorname{Re} \int_{-\infty}^{\infty} \frac{d \omega_{1}}{\omega_{1}}\left[1-\underbrace{F\left(\frac{m_{c}^{2}-i \varepsilon}{m_{b} \omega_{1}}\right)}_{\text {perturbative }}+\frac{m_{b} \omega_{1}}{12 m_{c}^{2}}\right] \underbrace{h_{17}\left(\omega_{1}\right)}_{\text {non-perturbative }}.
\end{eqnarray}  
The perturbative component in $\Lambda_{17}$ arises from the uncut loop. The soft function $h_{17}$ cannot be extracted from data, so it is modeled from the known information of the moments of $h_{17}$. For example, the zeroth moment of $h_{17}$ is related to the mass difference of $B$ and $B^*$ \cite{Benzke:2010js}. Recently, there have been new developments towards obtaining the information of moments \cite{Mannel:2010wj, Gambino:2016jkc}. With this new knowledge regarding the moments of $h_{17}$ a new model for $h_{17}$ can be developed to better constrain the $Q_{1}^q-Q_{7\gamma}$ contribution. The first question is how to extract these new moments? 
\section{Moments of $h_{7}$}
The function $h_{17}$ can be thought of as the gluon PDF of a $B$ meson. It is given as a non local operator matrix element defined by
\begin{eqnarray}\label{h17}
h_{17}(\omega)=\int \frac{d r}{2 \pi} e^{-i \omega_{1} r} \frac{\left\langle\overline{B}\left|\left(\overline{h} S_{\overline{n}}\right)(0) \mu\left(1+\gamma_{5}\right) i \gamma^{\perp} \overline{n}_{\beta}\left(S_{\overline{n}} g G^{\alpha \beta} S_{\overline{n}}\right)(r \overline{n})\left(S_{\overline{n}}^{\dagger} h\right)(0)\right| \overline{B}\right\rangle}{2 M_{B}}
\end{eqnarray} 
where $S_{n(\bar{n})}$ are Wilson lines. Using integration by parts one can show that $k^{\text{th}}$ moment is given by 
\begin{eqnarray}\label{kthMoment}
\left\langle\omega_{1}^{k} h_{17}\right\rangle=\left.(-1)^{k} \frac{1}{2 M_{B}}\left\langle\overline{B}\left|\left(\overline{h} S_{\overline{n}}\right)(0) \cdots(i \overline{n} \cdot \partial)^{k}\left(S_{\overline{n}}^{\dagger} g G_{s}^{\alpha \beta} S_{\overline{n}}\right)(r \overline{n})\left(S_{\overline{n}}^{\dagger} h\right)(0)\right| \overline{B}\right\rangle\right|_{r=0},
\end{eqnarray}
where $igG_{s}^{\alpha \beta}=\left[iD^{\alpha},iD^{\beta}\right]$. The expression $(i \overline{n} \cdot \partial)^{k}\left(S_{\overline{n}}^{\dagger} g G_{s}^{\alpha \beta} S_{\overline{n}}\right)$ in equation (\ref{kthMoment}) can be evaluated using the following identity \cite{Gunawardana:2019gep}
\begin{eqnarray}\label{newIdentity}
i \overline{n} \cdot \partial\left(S_{\overline{n}}^{\dagger}(x) O(x) S_{\overline{n}}(x)\right)=S_{\overline{n}}^{\dagger}(x)[i \overline{n} \cdot D, O(x)] S_{\overline{n}}(x),
\end{eqnarray}
which transforms derivatives to the commutators of covarient derivatives. Therefore, $k$ derivatives in equation (\ref{kthMoment}) convert into $k$ number of commutators. From this we obtained the general result
\begin{eqnarray}
\langle\omega_{1}^{k} h_{17}\rangle=(-1)^k\frac{1}{2M_B}\langle\bar{B}|\bar{h}\slashed{\bar{n}}(1+\gamma_5)\gamma_{\alpha}^{\perp}\underbrace{[i\bar{n}\cdot D,[i\bar{n}\cdot D,\cdots[i\bar{n}\cdot D,}_{k\,\text{times}}[D^{\alpha},i\bar{n}\cdot D]\cdots]]s^{\lambda}h|\bar{B}\rangle\nonumber.
\end{eqnarray}
In \cite{Gunawardana:2017zix} we provided a systematic approach to relate such matrix elements to commutators of HQET parameters. From this we found the following results for the first two moments
\begin{eqnarray}
\begin{array}{l}{\left\langle\omega_1^0 h_{17}\right\rangle= 2 \lambda_{2}=2 \mu_{G}^{2} / 3} \\ {\left\langle\omega_{1}^{2} h_{17}\right\rangle=\frac{2}{15}\left(5 m_{5}+3 m_{6}-2 m_{9}\right)},\end{array}
\end{eqnarray}
where $\mu_{G}^2$ and $m_{i}$ are non perturbative parameters that can be obtained from the data. The numerical values for $m_i$ were first obtained in \cite{Gambino:2016jkc}, which allowed us to obtain the new numerical estimates of these moments of $h_{17}$. Following this, the values of $\langle\omega_1^0h_{17}\rangle$ and $\langle\omega_1^2h_{17}\rangle$ are \cite{Gunawardana:2019gep}
\begin{eqnarray}\label{NumericalEstimates}
\begin{aligned}\left\langle\omega_{1}^{0} h_{17}\right\rangle &= 0.237 \pm 0.040 \mathrm{\,GeV}^{2} \\\left\langle \omega_{1}^{2} h_{17}\right\rangle &= 0.15 \pm 0.12 \mathrm{\,GeV}^{4}. \end{aligned}
\end{eqnarray}
Although the numerical error on the moments are large compered to the central values, they still give important information. For example, the model found in \cite{Benzke:2010js} predicts  $\langle\omega_1^2 h_{17}\rangle\in\left[-0.31,0.49\right]\text{ GeV}^4$. Whereas, using equation (\ref{NumericalEstimates}), $\left\langle \omega_{1}^{2} h_{17}\right\rangle \in \left[0.03,0.27\right]\text{ GeV}^4$. Therefore, our estimates are significantly smaller in range compared to the 2010 model estimates. In future, the data from Belle $II$ and lattice QCD (LQCD) can further improve these estimates.
\section{Applications}
\subsection{New model for $h_{17}$}
The function $h_{17}$ defined in equation (\ref{h17}) has the following properties: it is a real and even function over gluon momentum $\omega_1$, it's odd moments over $\omega_1$ vanish and it has dimensions of mass.  Our model was built by using the combination of \textit{Hermite polynomials} and a \textit{Gaussian} with width $\sigma$. Thus \cite{Gunawardana:2019gep}
\begin{eqnarray}\label{Model}
h_{17}(\omega_1)=\sum_{n}a_{2n}H_{2n}(\frac{\omega_1}{\sqrt{2}\sigma})e^{\frac{-\omega_1^2}{2\sigma}},
\end{eqnarray}
where the coefficients $a_{2n}$ are related to the moments as follows
\begin{eqnarray}
{a_0}=\frac{\left\langle\omega_{1}^{0} h_{17}\right\rangle}{\sqrt{2 \pi}|\sigma|}, \quad a_{2}=\frac{\left\langle\omega_{1}^{2} h_{17}\right\rangle-\sigma^{2}\left\langle\omega_{1}^{0} h_{17}\right\rangle}{4 \sqrt{2 \pi}|\sigma|^{3}}, \quad a_{4}=\cdots\nonumber.\end{eqnarray} 
Even though we only have numerical estimates up to $\langle\omega_1^2h_{17}\rangle$, we assumed the conservative bounds $\langle\omega_1^4h_{17}\rangle\in\left[-0.3,0.3\right]\text{ GeV}^6$ and $\langle\omega_1^6h_{17}\rangle\in \left[-0.3,0.3\right]\text{ GeV}^8$ to evaluate the non-perturbative error. Since the $h_{17}$ is a soft function, we further constrain $h_{17}$ so that $|h_{17}(\omega_1)|\leq 1\text{ GeV}$ and require it to have no structures beyond $|\omega_1|\geq 1\text{ GeV}$. Scanning through all possible numerical values of moments up to $\langle\omega_1^6h_{17}\rangle$ we found $\Lambda_{17}\in\left[-24,5\right]\text{ MeV}$. See \cite{Gunawardana:2019gep} for further details.
\subsection{CP asymmetry}\label{CPasymmetry}
The SM prediction for the CP asymmetry comes only from $Q_{1}-Q_{7\gamma}$ contribution, which is dominated by the non-perturbative effects from resolved photon contributions. The resulting SM prediction is $-0.6 \%<\mathcal{A}_{X_{s} \gamma}^{\mathrm{SM}}<2.8 \%$ \cite{Benzke:2010tq}. This estimate can be compared to the 2019 update of the 2018 PDG experimental value of $1.5 \% \pm 1.1 \%$ \cite{Tanabashi:2018oca}.\par
The $Q_{1}^q-Q_{7\gamma}$ contribution to the CP asymmetry $\mathcal{A}_{X_{s} \gamma}^{\mathrm{res}, 17}$ is
\begin{eqnarray}
\mathcal{A}_{X_{s} \gamma}^{\mathrm{res}, 17}=\frac{\pi}{m_{b}}\left\{\operatorname{Im}\left[\left(1+\epsilon_{s}\right) \frac{C_{1}}{C_{7 \gamma}}\right] \tilde{\Lambda}_{17}^{c}-\operatorname{Im}\left[\epsilon_{s} \frac{C_{1}}{C_{7 \gamma}}\right] \tilde{\Lambda}_{17}^{u}\right\}
\end{eqnarray}    
where
\begin{eqnarray}
\tilde{\Lambda}_{17}^{u}=\frac{2}{3} h_{17}(0), \quad \tilde{\Lambda}_{17}^{c}=\frac{2}{3} \int_{4 m_{c}^{2} / m_{b}}^{\infty} \frac{d \omega}{\omega} f\left(\frac{m_{c}^{2}}{m_{b} \omega}\right) h_{17}(\omega)
\end{eqnarray} 
with 
\begin{eqnarray}
f(x)=2 x \ln \frac{1+\sqrt{1-4 x}}{1-\sqrt{1-4 x}}.
\end{eqnarray}
In 2010 estimates for the parameters $\Lambda_{17}^u$ and $\Lambda_{17}^c$ were given as $-330 \mathrm{\,MeV}<\tilde{\Lambda}_{17}^{u}<+525 \mathrm{\, MeV}$ and $-9 \mathrm{\, MeV}<\tilde{\Lambda}_{17}^{c}<+11 \mathrm{\,MeV}$ \cite{Benzke:2010tq}. The new model for $h_{17}$ allows us to reevaluate these estimates.\par
Because of the constraint $\left|h_{17}\left(\omega_{1}, \mu\right)\right| \leq 1 \mathrm{\,GeV}$, the smallest and the largest values that $h_{17}(0)$ can have is $\pm 1 \text{ GeV}$. In the two Hermite polynomial model we can change the value of the parameter $\sigma$ in equation (\ref{Model}) such that the $|h_{17}(0)|=1\text{ GeV}$ for the given set of moments. For example, the central values of $\langle\omega_1^0 h_{17}\rangle$ and $\langle\omega_1^2 h_{17}\rangle$ gives $h_{17}(0)=-1\text{ GeV}$ for $\sigma=0.27\text{ GeV}$. By considering the moments up to $\langle\omega_1^4\rangle$, we found $h_{17}(0)=+1\text{ GeV}$. See \cite{Gunawardana:2019gep} for further details. Therefore, this provides a new range $\Lambda_{17}^u\in\left[-660,660\right]\text{ MeV}$.\par
Scanning through the different $\sigma$ and moment values upto $\langle\omega_1^6h_{17}\rangle$ we found the new estimate for $\tilde{\Lambda}_{17}^{c}$ as $\tilde{\Lambda}_{17}^{c} \in[-7,10] \text { MeV }$ \cite{Gunawardana:2019gep}. 
\section{Phenomenological estimates } 
Based on the new estimates of $\Lambda_{17},\tilde{\Lambda}_{17}^u$ and $\tilde{\Lambda}_{17}^c$ we update the results found in \cite{Benzke:2010js}  and \cite{Benzke:2010tq}. The $Q_{1}^q-Q_{7\gamma}$ contribution to the total uncertainty was evaluated by using $C_{1}(\mu)=1.257,C_{7}(\mu)=-0.407$ (calculated at $\mu=1.5\text{ GeV}$) and $m_b=4.58\text{ GeV}$. This gives
\begin{eqnarray}
\left.\mathcal{F}_{E}\right|_{17} \in[-0.3,+1.6] \%
\end{eqnarray} 
This new range should be compared to the 2010 range $\left[-1.7,+4.0\right]\%$ found in \cite{Benzke:2010js}. The total uncertainty of the rate can be obtained by using $\left.\mathcal{F}_{E}\right|_{88} \in[-0.3,+1.9] \%$ \cite{Benzke:2010js} along with either $\left.\mathcal{F}_{E}\right|_{78} ^{\mathrm{VIA}} \in[-2.8,-0.3] \%$ or the new experimental value from PDG, $\left.\mathcal{F}_{E}\right|_{78} ^{\exp } \in[-1.4,+2] \%$. Scanning over various contributions give
\begin{eqnarray}
-3.4 \%<\mathcal{F}_{E}(\Delta)<+3.2 \% \quad(\text { using } \mathrm{VIA})
\end{eqnarray}
This new range should be compared with the 2010 range $-4.8 \%<\mathcal{F}_{E}(\Delta)<+5.6 \%(\text { using } \mathrm{VIA})$ \cite{Benzke:2010tq}, and it implies a reduction to the total error by a third. In contrast, by using the experimental estimate the new range becomes  
\begin{eqnarray}
-2.0 \%<\mathcal{F}_{E}(\Delta)<+5.5 \%\quad \text{ using exp.}
\end{eqnarray}  
Compared to the 2010 range $-6.4 \%<\mathcal{F}_{E}(\Delta)<+11.5 \%\left(\text { using exp.}  \right)$ \cite{Benzke:2010js}, the new estimate reduces the total error by a half.\par
Plugging in our new estimates for $\tilde{\Lambda}_{17}^u$ and $\tilde{\Lambda}_{17}^c$ found in the section \ref{CPasymmetry} to the following expression
\begin{eqnarray}
\mathcal{A}_{X_{s} \gamma}^{\mathrm{SM}}=\left(1.15 \times \frac{\tilde{\Lambda}_{17}^{u}-\tilde{\Lambda}_{17}^{c}}{300 \mathrm{MeV}}+0.71\right) \%
\end{eqnarray}
gives us $-1.9 \%<\mathcal{A}_{X_{s} \gamma}^{\mathrm{SM}}<3.3 \%$. This should be compared to the 2010 range $-0.6 \%<\mathcal{A}_{X_{s} \gamma}^{\mathrm{SM}}<2.8 \%$ in \cite{Benzke:2010tq}. 
\section{Conclusions and outlook}
The radiative decay $\bar{B}\rightarrow X_s\gamma$ is an important new physics probe. However, its standard model prediction of the total rate contains a $7\%$ uncertainty. Non-perturbative $1/m_b$ effects dominates this with $5\%$ contribution. These $1/m_b$ effects are obtained from the operator pairs $Q_1^q-Q_{7\gamma}, Q_{8g}-Q_{8g}$ and $Q_{7\gamma}-Q_{8g}$. They are parameterized by a perturbatively calculable part and a non-perturbative function denoted by $h_{17}$. Currently, with new Belle data on $Q_{7\gamma}-Q_{8g}$, the largest contributor to the uncertainty is from $Q_1^q-Q_{7\gamma}$. The recent progress in our knowledge on moments of $h_{17}$ allows us to extract them numerically. Therefore, a new model based on the higher order moments of $h_{17}$, which can incorporate also future information on moments, was developed \cite{Gunawardana:2019gep} to reduce this uncertainty.\par
We found the two lowest moments of soft function $h_{17}$ using data given in \cite{Gambino:2016jkc} and the method developed in \cite{Gunawardana:2017zix}. Since $h_{17}$ is a even function over the gluon momentum and has the dimensions of mass, we developed a new systematic model based on a combination of \textit{Hermite polynomials} and a \textit{Gaussan}. The explicit form of the new model is given in (\ref{Model}). Also, further constraints were employed such as $\left|h_{17}\left(\omega_{1}\right)\right| \leq 1\mathrm{\, GeV}$ and limit the function from having structures beyond $|\omega_1|\geq 1\text{ GeV}$. Scanning through different values of moments we found a new estimate for $Q_1^q-Q_{7\gamma}$. Our estimate reduced 2010 estimate \cite{Benzke:2010js} by a third. Also, a new range for total rate was obtained by combining the new estimate for $Q_{7\gamma}-Q_{8g}$ with our new result for $Q_1^q-Q_{7\gamma}$. It is reduced by half compared to the 2010 values \cite{Benzke:2010js}.\par
The SM prediction for CP asymmetry is obtained by non-perturbarive parameters $\tilde{\Lambda}_{17}^u$ and $\tilde{\Lambda}_{17}^c$. These parameters are also related to $h_{17}$. We reevaluated their ranges using our analysis. From this we found a new estimate for SM CP asymmetry as $-1.9 \%<\mathcal{A}_{X_{s} \gamma}^{\mathrm{SM}}<3.3 \%$, which is an increased range compared to the 2010 estimate \cite{Benzke:2010tq}. This is because of the increased range of the $\tilde{\Lambda}_{17}^u$.\par
We conclude our discussion with a remark on future developments. With the new information on the moments we can better control the hadronic effects. However, the scale dependence on $1/m_b$ corrections are not fully controlled because currently we treat them at the leading order in $\alpha_s$. Therefore, to improve the $Q_{1}^q-Q_{7\gamma}$ contributions further, we need to take account of the $\alpha_s$ corrections.\par
Our model relies on the numerical estimates of the matrix elements of dimension 8 operators, but it could further improved if we knew the  numerical estimates of dimension 9 matrix elements. With the Belle II data we can hope to have improvements on this.\par
Finally, in this discussion we considered the quantities that are integrated over photon energy. The above moment information can be used to model $Q_{1}^q-Q_{7\gamma}$ contributions for quantities that are not integrated over photon spectrum. However, this is left for future work.
\section{Acknowledgment}
This work was supported by U.S. Department of Energy grant DE-SC0007983.

\end{document}